\documentclass[letterpaper, 10 pt, conference]{ieeeconf}
\IEEEoverridecommandlockouts
\usepackage{amsmath}
\usepackage{amssymb}
\usepackage{amsfonts}
\usepackage{latexsym}
\usepackage{color}
\usepackage{verbatim}
\usepackage{xspace}
\usepackage{float}
\usepackage[linesnumbered,ruled,vlined]{algorithm2e}
\usepackage{algpseudocode}
\usepackage{url}
\usepackage{graphicx}
\usepackage{caption}
\usepackage{subcaption}
\usepackage{graphicx}
\usepackage{blindtext}
\usepackage[noadjust]{cite}
\usepackage{multicol}
\usepackage{amsmath}
\usepackage{graphicx}
\usepackage[colorinlistoftodos]{todonotes}
\usepackage[colorlinks=true, allcolors=blue]{hyperref}

\usepackage{tikz}
\newcommand{\Cross}{\mathbin{\tikz [x=1.2ex,y=1.2ex,line width=.1ex] \draw (0,0) -- (1,1) (0,1) -- (1,0);}}%

\SetKwInput{kwInit}{Initialization}
\newtheorem{definition}{Definition}[section]
\newtheorem{assumption}{Assumption}[section]

\renewcommand{\thefigure}{\arabic{figure} }

\usepackage[left=0.75in, right=0.75in,top=0.75in ,bottom=0.75in]{geometry}
\begin{document}

\thispagestyle{empty}
\author{%
	Pooyan Ehsani$^{1}$ \and Jia Yuan Yu$^{1}\footnotemark$
	 \thanks{P. Ehsani and J. Y. Yu, are with the Concordia Institute for Information Systems Engineering, Concordia University
		Montreal, Canada.	
	Email: {\tt\small p\_ehsan@encs.concordia.ca, jiayuan.yu@concordia.ca}}
}
\title{\vspace{0.25in}The Merits of Sharing a Ride}

\maketitle
\begin{abstract}
The culture of sharing instead of ownership is sharply increasing in individuals behaviors. Particularly in transportation, concepts of sharing a ride in either carpooling or ridesharing have been recently adopted. An efficient optimization approach to match passengers in real-time is the core of any ridesharing system. In this paper, we model ridesharing as an online matching problem on general graphs such that passengers do not drive private cars and use shared taxis. We propose an optimization algorithm to solve it. The outlined algorithm calculates the optimal waiting time when a passenger arrives. This leads to a matching with minimal overall overheads while maximizing the number of partnerships. To evaluate the behavior of our algorithm, we used NYC taxi real-life data set. Results represent a substantial reduction in overall overheads.
\end{abstract}


% \fi
\IEEEpeerreviewmaketitle

\section{Introduction}
Currently, cars are vastly underutilized. Firstly, most cars sit in parking spaces most of the time. Secondly, cars underused due to their low occupancy rate relative to the number of seats available (1.67 persons per vehicle in the U.S. \cite{di2015exploration}). This inefficient use of transportation and parking resources had to be tolerated to enjoy reliable, safe, and on-demand transportation. Today,
however, technology allows us to move away from the car-ownership paradigm toward a shared transportation one \cite{gargiulo2015dynamic}, \cite{saranow2006carpooling}. Individuals who have similar itineraries and time schedules use a shared vehicle in any portion of their paths in a ridesharing scheme. An efficient use of empty seats of cars in urban trips has numerous economic and societal outcomes for individuals and the general public. Increasing warnings about high consumption rates of finite sources of fossil fuels and global warming, getting stuck at traffic congestions \cite{xiang2008study} and rising gas prices are few examples demanding wiser adoption of vehicles. Additionally, Sharing a car enables passengers to share travel expenses which could be a substantial source of motivation. It should be noted that ridesharing is different from conventional carpooling and vanpooling systems dedicated to trips on a regular basis \cite{levofsky2001organized} which are very limited with respect to scheduling.   

Taxi service providers can benefit from the advantages of ridesharing as well. These companies may lose some of the passengers due to their limited fleet size compared to the requests in metropolitan areas. Relatively high costs of taking taxis in comparison with other public transportation means is a deterrent factor of extensive usage of taxis. Having more than one passenger in a taxi could be a key factor in solving issues above since the occupancy rate of taxis increase, and every passenger pays marginally lower expenses.

In this paper, we focus on the case when a central coordinator is owning a finite fleet of cars offering an automated matching of passengers within an urban area. Optimal time windows and other constraints on accepted detours for every passenger need to be respected while overall costs minimized. Each active individual has a specific itinerary and has to make a detour and extra stops to accommodate ridesharing partner. The length of the individual optimal waiting time depends on the passenger's willingness to deviate from the shortest path to get to his destination.

In this paper, we design and implement an algorithm to maximize the matching of passengers (based on the matching algorithms on general graphs \cite{blossommod1965}) while optimizing the costs incurred for a taxi service provider. We evaluate the effectiveness of the proposed algorithm using real data set.  

We can formulate this problem in the sense of a cooperative agent based system \cite{xing2009smize}. In this case, neither of tasks can be carried out by individual agents in isolation. Thus, problem solving necessitates agents' interaction. User agents inquire an appropriate number of service agents to perform the matching task. Service agents maintain and has access to active passengers' information including origin, destination, and flexibility. To be precise, matching agents find a compatible match for a passenger willing to leave.

The remainder of this paper is organized as follows. In section \ref{Section 2}, formulates the model to solve online ridesharing problem and presents our proposed algorithm. Experimental results are discussed in Section \ref{section 3}. In Section \ref{Section 4}, we review related work and represent how our approach is different from others. Finally, we summarize conclusions and describe future works in Section \ref{section 5}.
\section{Problem Definition}\label{Section 2}
We model the ridesharing problem as an online single decision maker problem. The road network modeled as a weighted graph, with vertices representing locations, and weighted edges representing the distance between locations. Passengers appear at random time instants, with random source and destination and we fix a flexibility factor which models the overhead in travel time. We assume that a central decision maker keeps track of this information for each passenger in real-time, and is tasked with matching passengers in an online fashion.
For simplicity, we first assume that each passenger is either matched to one other passenger or travels alone. Moreover, once a trip started, matches are no longer allowed. In this model, the decision is made of two components: how long to make each passenger wait, and which another one (if any) to match to when the waiting time has elapsed. Once passengers are matched, we assume that a nearby taxi is dispatched to fulfill the trip. A number of objectives can be considered: from a social welfare perspective. We want to minimize for each passenger the expected distance traveled in single occupancy, subject to the flexibility constraint. From the perspective of an operator like UberPool, the goal is to maximize the expected profit for each passenger. We propose to estimate the distribution of the random arrival time, source and  destination from openly available data. Given this distribution and travel times on the road network, we propose an online algorithm that first optimizes the waiting time for each newly appeared passenger, and then matches a passenger whose waiting time has elapsed by using the minimum cost perfect matching algorithm.
To model this problem, we present a set of passengers $P = \{p_1,\ldots,p_N\}$ where $p_i = (s_i,d_i)$, $i \in \{1,\ldots,N\}$. Each passenger has a starting point $s_i$ and a destination point $d_i$. 
The central ride sharing system assigns either one or two passenger in a single car. In the former, the taxi picks up passenger $p_i$ and moves along the shortest path from his starting point $s_i$ to his desired destination $d_i$. Conversely, if a matching occurs, the taxi picks up passenger $i$, travels to lift passenger $j$ and drops them off in their respective destinations. The taxi always opts for the shortest path available between two points. We assume that starting points and destinations are i.i.d. random variables and the distributions are known. In addition, partial sums of exponential random variables $e_k$, $k \in \{1,\ldots,N\}$ represent time elapsed between the appearance of consecutive passengers. $t_i$ denotes passenger $p_i$'s time of arrival.
\begin{equation}
t_i = \sum\limits_{k=1}^i e_k.
\end{equation}

\begin{assumption}
	We assume all random variables are generated i.i.d. and that distribution functions are known.
\end{assumption}

Let $G=(V,E)$ be a $q$ by $q$ weighted grid representing a road network. This graph illustrates starting points and destinations of passengers currently in the system. Let $\pi_i$ denote the shortest path travel time for passenger $i$ to move from his starting point to his desired destination. We assume that each passenger $p_i$ is willing to incur at most $\epsilon$ fraction above the shortest path. Consequently, the set of allowable paths for every arrived $p_i$ is denoted:
\begin{align}
\Pi_i \triangleq \{\pi : |\pi|\leq (1 + \epsilon_i) |\pi_i|\}
\end{align}

The set $\Pi_i$ contains every allowable path from $s_i$ to $d_i$.  In this paper, we design an algorithm to find the optimal waiting time for every passenger $p_i$ arriving and then decide on the best matching to optimize the system-wide overall overhead.
The optimum waiting time for passenger $p_i$ before moving on is a function of $s_i, d_i$, and $\epsilon_i$ which is indicated by $\tau_i(s_i,d_i,\epsilon_i)$. For simplification, we denote this by $\tau_i$. We assume that cars always show up by the end of $\tau_i$ and no passenger waits longer because a taxi arrives late. Different values of $\tau_i$ may yield more compatible matches and consequently lower overheads. A passenger $p_i$ spends a longer time  before reaching his destination as a result of participating in a ride sharing arrangement. By denoting actual travel time as $\omega_i$ the following inequality must hold for any passenger entering the system. 
\begin{align}\label{enq}
\tau_i + \omega_i \leq (1+\epsilon_i) \pi_i
\end{align}
As we increase the waiting time, more passengers appear in the pool of
waiting passengers; however, as the waiting time for a given passenger increases, the tolerable detour distance for that passenger decreases, such that fewer matches within the pool are feasible due to the flexibility constraint.
We define set $W_t$ including all passengers who are still in the system at any instance $t$. This set excludes matched up passengers who have left the system before $t$.
\begin{align}
W_t = \{p_i \ | \ t_i \leq t < t_i + \tau_i \} \ , \ i \in \{1,\ldots,N\}
\end{align}
During $\tau_i$, some passengers may enter the system, and some of them proper for share a ride with $p_i$ based on their source and destination. These nominated passengers add to set $Q_i \subset W$ which is initially null for every passenger $p_i$. By the end of $\tau_i$, the central ride sharing system decides to assign $p_i$ to the best match to optimize system-wide overhead. 

\begin{definition}
(Compatible Matching). A match between $p_i$ and $p_j$ are compatible if there exist two paths $\pi_i \in \Pi_i$ and $\pi_j \in \Pi_j$ which:
\begin{align}
	d_i, s_j \in V(\pi_i \cap \pi_j) \ \textrm{or} \ s_j, d_j\in V(\pi_i \cap \pi_j) \nonumber\\
	\textrm{or} \ s_i, d_i\in V(\pi_i \cap \pi_j)  \ \textrm{or} \ d_j, s_i \in V(\pi_i \cap \pi_j)
\end{align}
\end{definition}

\begin{definition}
	(Order of sharing). Define $O(p_i,p_j)$ as the cost associated with the order of pick up and drop off of the passengers such that the first element in the pair represents the first passenger picked up and the second element denotes the first passenger dropped off in his desired destination. It should be noted that if $p_i$ does not share a ride, then the cost is simply the cost of going alone on the shortest path, denoted by $O(p_i,0)$.
\end{definition}

\begin{definition}
	(Cost of a matched pair). Define the travel cost for matched up passengers $p_i$ and $p_j$, $j \in Q$:

	\begin{align}
	C(p_i,p_j) = 
	\begin{cases}
	\displaystyle \frac{1}{2}\min_{m,n \in \{i,j\}} \big\{O(p_m,p_n)  \big\} &\mbox{if } i \neq j \\
	O(p_i,0) &\mbox{if } i = j  
	\end{cases}
	\end{align}
\end{definition}

It should be noted that the cost is proportional to the distance covered by vehicles like fuel cost. We assume that total overhead is comprised of two parts, total waiting overhead and total travel overhead. As described before, an increase in waiting time has a limiting effect on feasible ride sharing partners for passenger $i$; therefore, influences the probability of finding a match. The optimal waiting time (in the expected sense) for every passenger $i$ is as follows:
\begin{align}\label{tau}
\tau_i = &\arg\min\limits_{u \in [1,\infty)} \Psi_i (u) + \Gamma_i (u)
\end{align}
In which $\Psi_i (u)$ expectation shows the case which $p_j=\{s_j,d_j\} \in V \Cross V$  is not matched to passenger $p_i$ and is defined as follows:
\begin{align}
\Psi_i &\triangleq  \mathbb{E} \Big\{ \omega_i \, | \, \forall p_j \in V \Cross V: \, \Big(\textrm{$p_i$ and $p_j$ are not compatible} \Big) \Big\} \nonumber\\
&= \sum\limits_{p_j \in V \Cross V} \Big\{\mathbb{P} \Big(	d_i, s_j \notin \pi_i \cap \pi_j \ \textrm{and} \ s_j, d_j\notin \pi_i \cap \pi_j \nonumber\\
&\qquad\qquad\quad\textrm{and} \ s_i, d_i\notin \pi_i \cap \pi_j  \ \textrm{and} \ d_j, s_i \notin \pi_i \cap \pi_j \Big) \nonumber
 \nonumber\\&\quad\qquad\qquad C(p_i,p_i)
 \nonumber\\&\quad\quad\ +
\mathbb{P}  \Big(d_i, s_j \in \pi_i \cap \pi_j \ \textrm{or} \ s_j, d_j\in \pi_i \cap \pi_j \nonumber&\\
&\qquad\qquad\quad\textrm{or} \ s_i, d_i\in \pi_i \cap \pi_j  \ \textrm{or} \ d_j, s_i \in \pi_i \cap \pi_j \Big) \nonumber. \\& \qquad\qquad\ \  \Big(1-\mathbb{P}\big(t_i\leq \boldsymbol{t_j} \leq  t_i +u\big)\Big) C(p_i,p_i)	\Big\}
\end{align}
On the other hand, $\Gamma_i$ represents a case where a matching can occur between $p_i$ and $p_j = \{s_j,d_j\} \in V \Cross V$  . The expected value is shown in:
\begin{align}
\Gamma_i &\triangleq \mathbb{E} \Big\{ \omega_i(Q_i) \, | \, \forall p_j \in V \Cross V: \, \Big( \textrm{$p_i$ and $p_j$ are compatible} \Big) \Big\} \nonumber\\
&={}\sum\limits_{p_j \in V \Cross V} 
\Big\{\mathbb{P} \Big(d_i, s_j \in \pi_i \cap \pi_j \ \textrm{or} \ s_j, d_j\in \pi_i \cap \pi_j \nonumber&\\
&\qquad\qquad\quad\textrm{or} \ s_i, d_i\in \pi_i \cap \pi_j  \ \textrm{or} \ d_j, s_i \in \pi_i \cap \pi_j.\Big) . \nonumber\\&\qquad\qquad \ \ \    \mathbb{P} \big(t_i\leq \boldsymbol{t_j} \leq  t_i +u\big)C(p_i,p_j) \Big\}
\end{align}
We assumed passengers arrival time are independent and identically distributed exponential random variables, $X \sim exp(\lambda)$ and $Y \sim exp(\lambda)$. We name the random variable $Z = X - Y$ and we know the CDF: $\mathbb{P}(X-Y \leq u)= 1 - \frac{e^{-\lambda u}}{2}$
By knowing the distribution of random variables $s_j$ and $d_j$, the probabilities in the above equation are determined. We consider a case where the travel cost is divided evenly between pairs in any arrangement.
Equation \ref{tau} prevent the algorithm to assign long waiting time to passenger.as the waiting time becomes longer, because of equation \ref{enq} must be held for passenger, the number of incompatible passengers increased, and it causes an increase in $\Psi_i (u)$.
\begin{definition}
	(Leaving candidates). Define a passenger leaving the system with the minimum amount the time of arrival and the optimum waiting time among all the current passengers.
	\begin{align}\label{candidate}
	k^*=\arg\min\limits_{\{i|p_i \in W\}}\{t_i + \tau_i\}\ \mbox{and} \  \theta =\min\limits_{\{i | p_i \in W\}} \{t_i + \tau_i\}
	\end{align}
	The number $k^*$ indicates the nominated passenger is leaving the system by the end of his optimal waiting time. $\theta$ denotes the latest departure time for the nominated passenger.. 
\end{definition}
\subsection{\textbf{Proposed Algorithm}}
In this section, we describe how our algorithm works to find the best matching over all passengers.
\begin{algorithm}[!h]
	\DontPrintSemicolon	
	\KwIn{
		
		- A graph $G$ and a positive number $N$\par
		- A set of pairs  $P = \{p_1,\ldots,p_N\}$, where $p_i = (s_i,d_i)$ and $\forall i \in \{1,\ldots,N\} : s_i, d_i \in V$\par
		- Pairs of $O(p_i,p_j)$ which is the cost associated with the order of pick up and drop off of passengers $i,j \in \{1,\ldots,N\}$
		- Distribution functions of $s_i, d_i, \epsilon_i, e_i$\par
		- A cost function $C : P^2 \rightarrow \mathbb{Z} \cup \{0\}$\par
	}
	\KwOut{
		
		- The optimal waiting time, i.e. value of $\tau_i$ for every passenger $p_i$\par
		- A matching $M \subseteq P^2$ such that $\forall i \in \{1,\ldots,N\} $ there is precisely one pair in $M$ such that $p_i$ appears in that pair (note that $c_i$ could be in both coordinates of the pair, in which case a passenger is paired with itself).
	}
	\medskip
	\kwInit{
		
	Create set of the current passengers such that $W_t = \{p_1\}$\;
	Evaluate optimal waiting time for the first passenger, $\tau_1$ as in equation \ref{tau}.\;
	Update the potential leaving candidate information, $k^*$, and $\theta$\ as in equation \ref{candidate}\; 
	$j \gets 2$\;
	
	\While{$|M| < N$}{
		\While{$\theta \geq t_j$}
		{
			Add the arrived passenger to the set of current ones in the system, $W_\theta \gets W_\theta + \{p_j\}$\;
			Evaluate the optimal waiting time for the newly arrived passenger,  $\tau_j$\;
			Update the leaving candidates and put it in set $k^*$\;
			Update the time of leaving for the candidate and tag it as $\theta$\;
			$j \gets j + 1$\;
		}
		Create a weighted graph $T$ with $|W_\theta|$ nodes and for each candidate add edge to compatible passengers in $W_\theta$ with a weight equal to the cost function $C$.\;
		Use Blossom algorithm to find minimum cost perfect matching in $T$.
		Remove matched up passengers from $W_\theta$ and add them as a pair to set $M$.\;
		Update the information about leaving candidate, $k^*$ and $\theta$\;
	}
}
	\caption{Finding the ride sharing matching}
	\label{alg:alg1}
\end{algorithm}

\section{Experimental Results}\label{section 3}
\begin{figure}[hbt!]
	\includegraphics[width=0.5\textwidth]{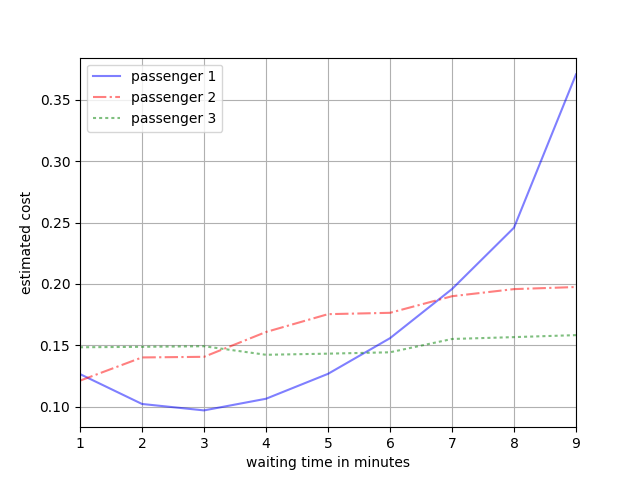} 
	\caption{Estimated $\Psi_i (u) + \Gamma_i (u)$}
	\label{Fig:passengers}
	\label{fig}
\end{figure}

\begin{figure}[hbt!]
	\centering
	\subcaptionbox{Original pickup locations  \label{fig:Scatter}}{%
		\includegraphics[width=0.45\columnwidth]{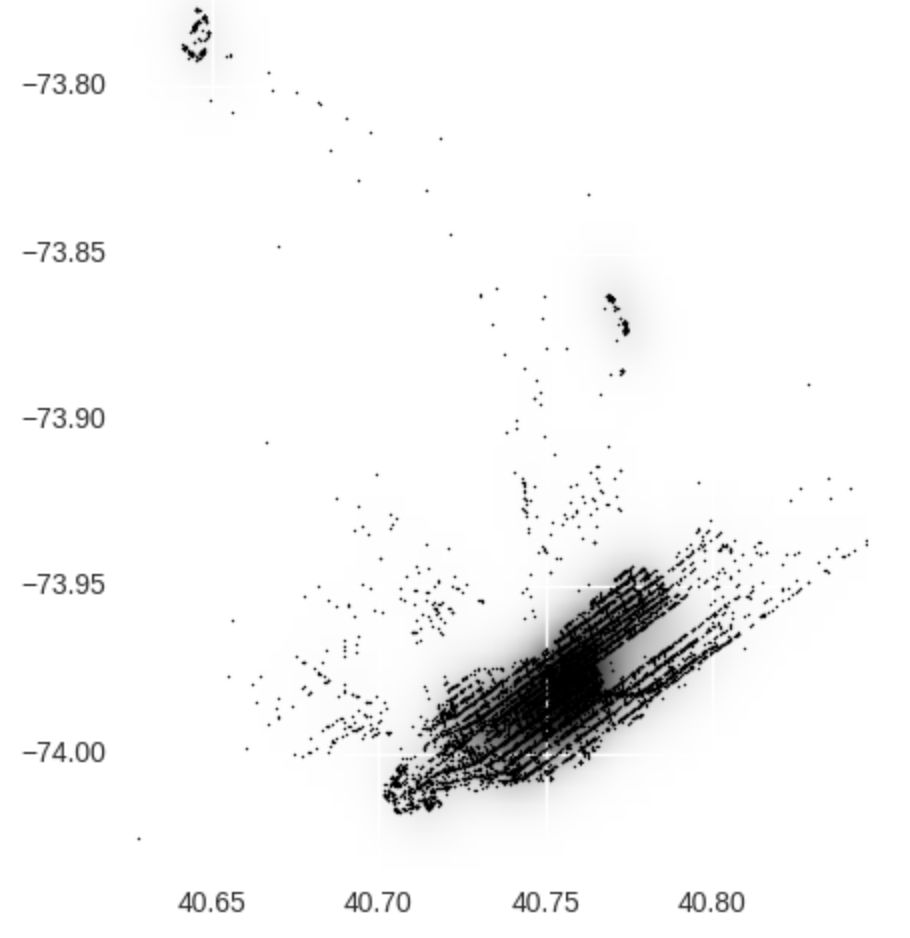}}
	\subcaptionbox{Estimated probability distribution function  \label{fig:KDE}}{%
		\includegraphics[width=0.45\columnwidth]{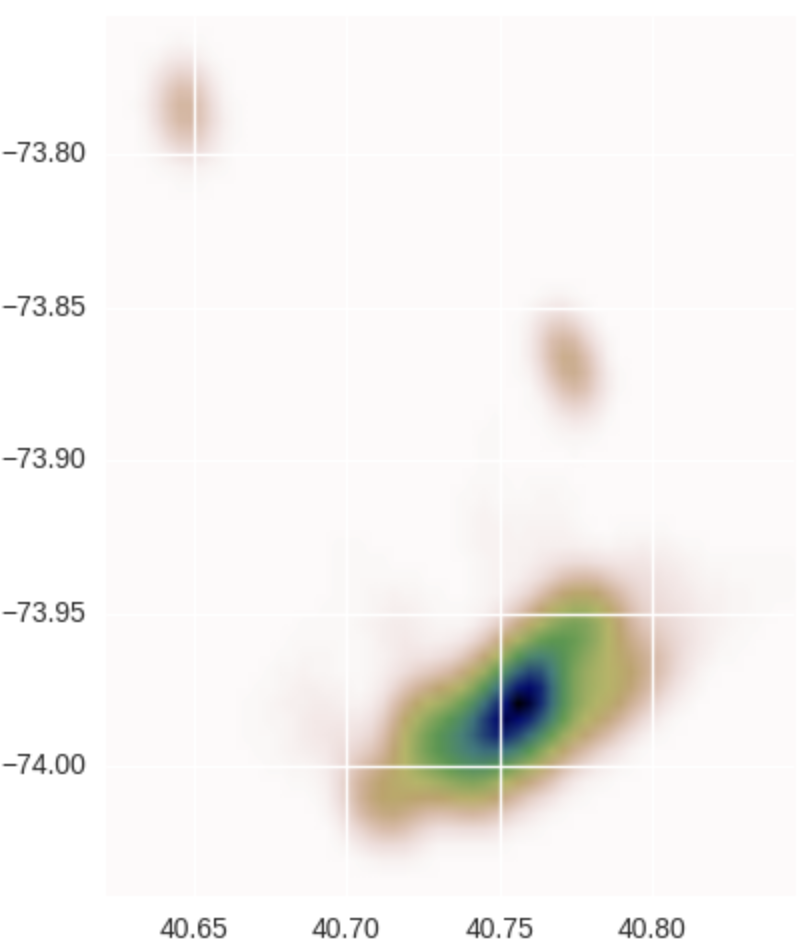} } \\
		\subcaptionbox{Original dropoff  locations  \label{fig:Scatter_D}}{%
		\includegraphics[width=0.45\columnwidth]{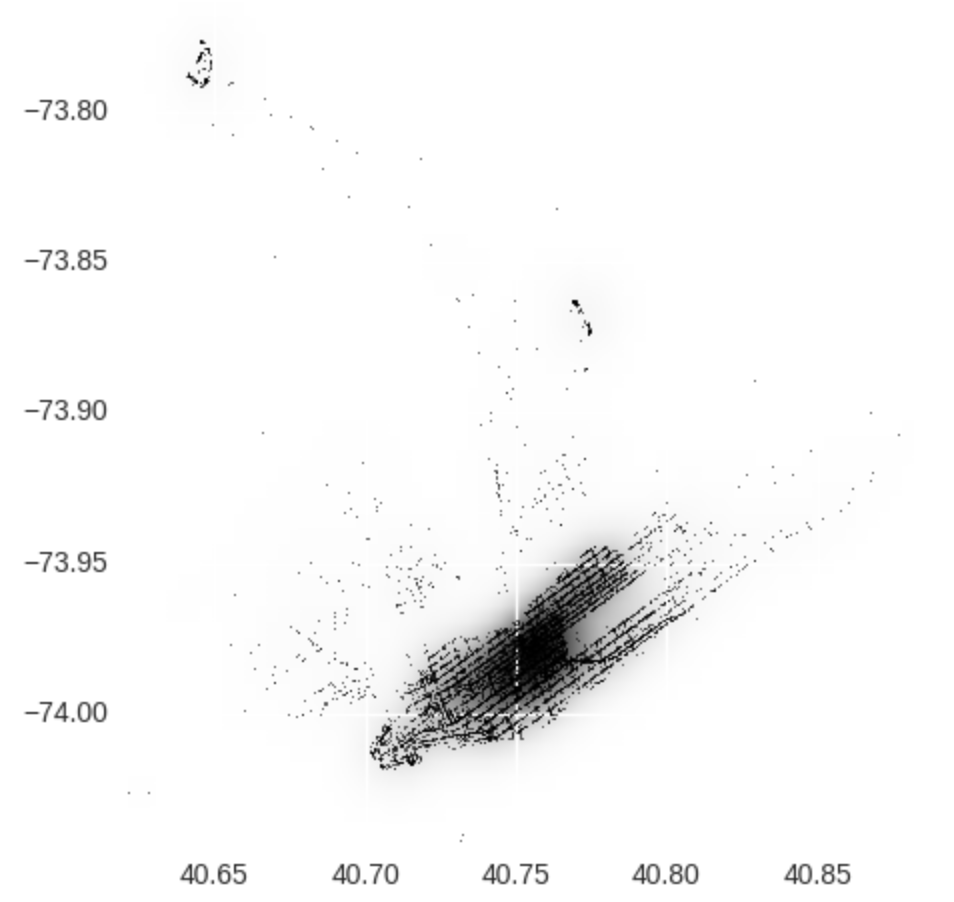}} 
	\subcaptionbox{Estimated probability distribution function  \label{fig:KDE_D}}{%
		\includegraphics[width=0.45\columnwidth]{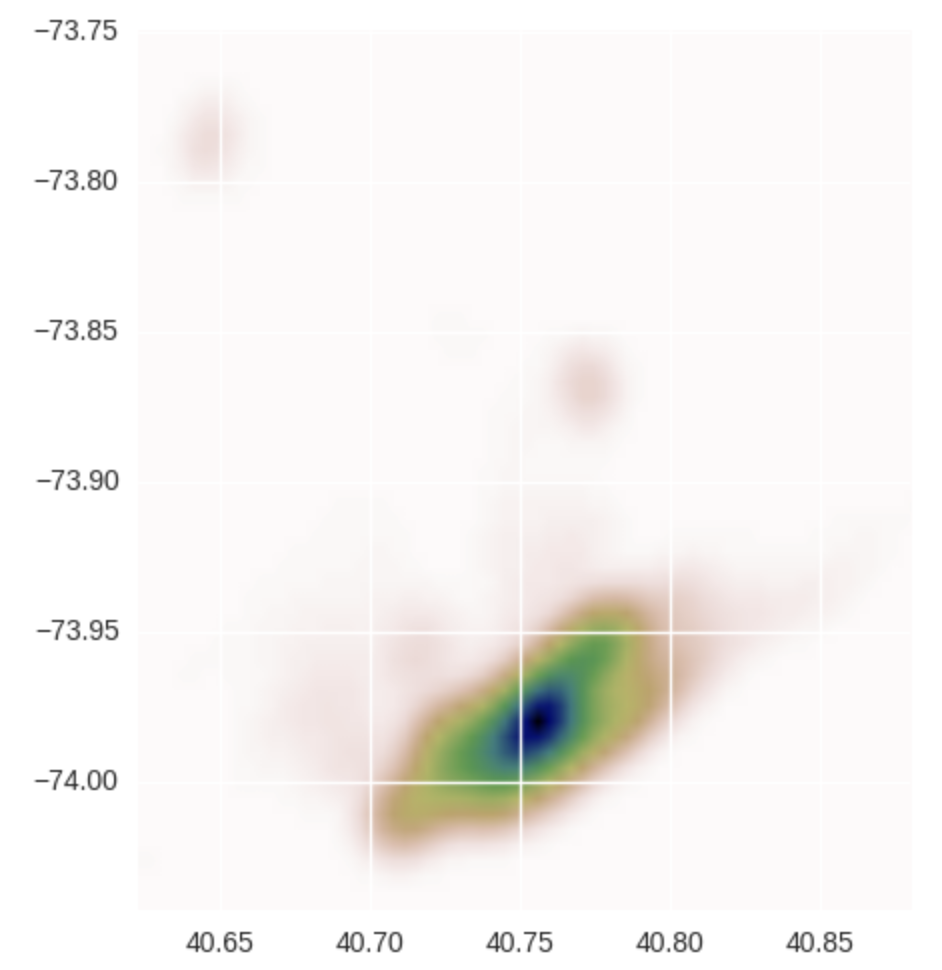} } \\
	\caption{Original nodes and estimated PDFs}\label{Fig:plot1}
\end{figure}

To test the behavior of our algorithm, we used the data provided by NYC Taxi \& Limousine Commission \cite{NYC} for yellow cabs in January 2016. This dataset contains more than 77 million trips information. Thus, we extracted 10,000 trips and used Kernel Density Estimation \cite{wand1993comparison} to estimate the probability distribution function of starting points and destinations of all passengers. Note that the database size of 10,000 entries is a practical case and closely resembles the current daily amount of matches in a city.
To extend the data set to our approach of arbitrary origin and destination locations, we applied the estimated PDF to estimate the likelihood of appearance of pickup and drop-off points of all passengers. For each starting point which we picked up randomly from the region with the estimated probability, a destination point is located in the same way.
Figure \ref{Fig:passengers}shows the likelihood function behavior for three different passengers in a short period which passengers can wait for a match. The function evaluated as discussed before in equation \ref{tau}.
Figure \ref{Fig:plot1}compares the original nodes locations on the left to the right showing a population density plot of NYC to support the validity of the perturbation.

\begin{figure}[hbt!]
	\includegraphics[width=0.5\textwidth]{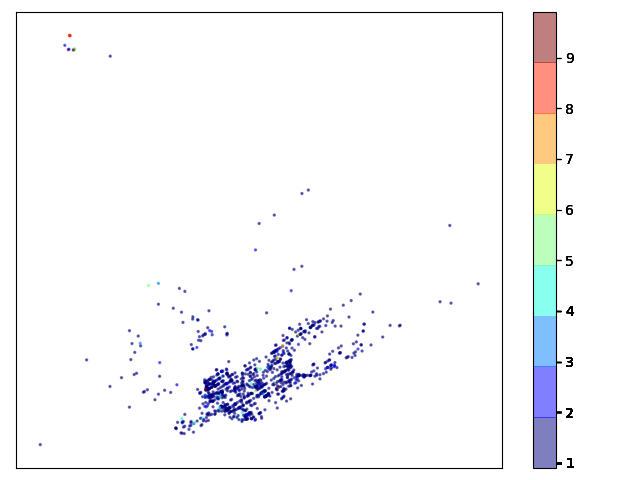} 
	\caption{Waiting time for passengers in minutes}
	\label{fig:subim3}
	\label{fig}
\end{figure}

\begin{figure}[hbt!]
	\includegraphics[width=0.5\textwidth]{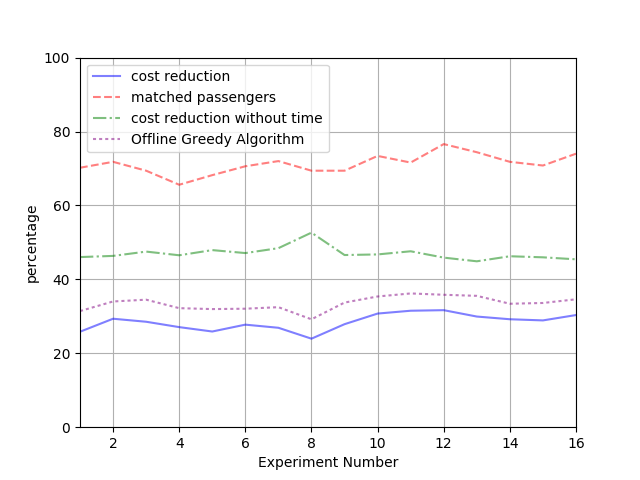} 
	\caption{The reduction in costs using ridesharing}
	\label{fig:subim4}
	\label{fig}
\end{figure}

\begin{figure}[hbt!]
	\includegraphics[width=0.5\textwidth]{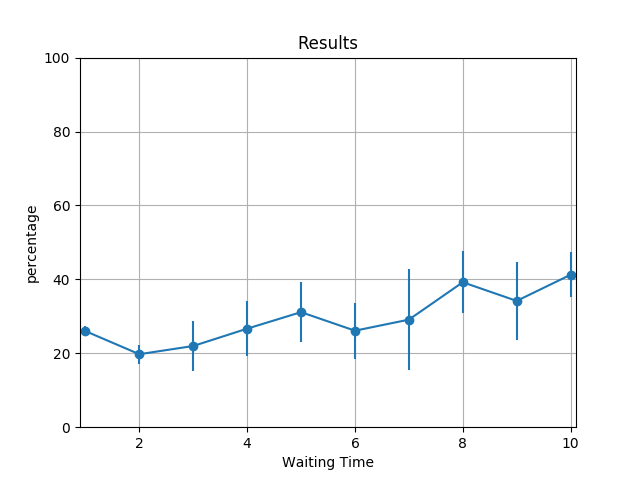} 
	\caption{The relation between waiting time and cost reduction}
	\label{fig:subim5}
	\label{fig}
\end{figure}

\begin{figure}[hbt!]
	\includegraphics[width=0.5\textwidth]{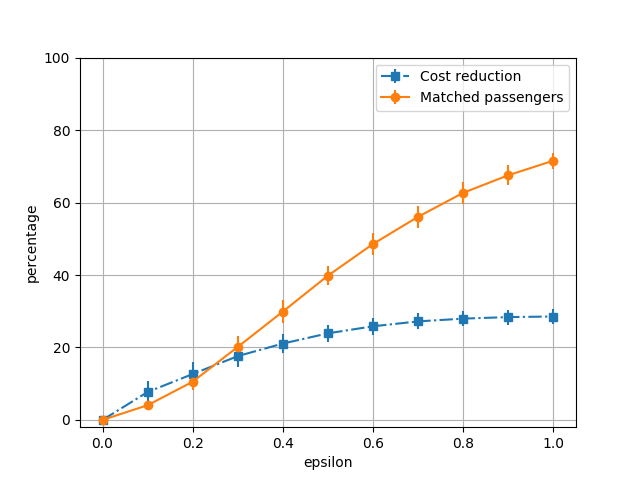} 
	\caption{The relation between $\epsilon$ and cost reduction}
	\label{fig:subim6}
	\label{fig}
\end{figure}

Our approach for estimating the waiting time applied for a set of 1000 passengers in March 2016 and as shown in figure \ref{fig:subim3} for most of the passengers  the waiting time is less than 2 minutes which is reasonable.
   Experiments on 16 datasets and each of them with the 1000 passengers from April shows that our algorithm has a significant reduction in costs using ridesharing. This plot shows that using ridesharing could result in a decrease in costs for a taxi provider company and its could motivate passengers to use their service more frequently. In \ref{fig:subim4}we compare cost reduction via our approach with a greedy offline algorithm which is for the greedy offline model, passengers arrival time are given and used to find the best waiting time from \ref{tau}. Blossom algorithm is used to find the proper matching of the passengers in the offline algorithm. For calculating the reduction in cost we define total travel distance of passengers as $\sum_{i=1}^{n} |\pi_i|$ where $n$ is the number of passengers in the dataset, and the total travel distance after finding the optimal matching as $M$ by algorithm \ref{alg:alg1} as $\sum_{e_i \in M} w_{e_i}$  and calculate the reduction in cost by $1- \dfrac{\sum_{e_i \in M} w_{e_i}}{\sum_{i=1}^{n} |\pi_i|}$.
In figure \ref{fig:subim5}we compare the reduction in costs on 16 test sets based on the waiting time of the passengers. As the waiting time increase for a specific passenger 
in figure \ref{fig:subim6}we increase the $\epsilon$ from zero to one in order to see how $\epsilon$ changes the results. We choose $\epsilon = 0.6$ based on experimental results.
By setting $\epsilon = 0.6$  the average waiting time for passengers based on 16 test sets equal to 1 minutes and 31 seconds and the average travel time increased by 3 minutes and 23 seconds.
\begin{figure}[hbt!]
	\includegraphics[width=0.5\textwidth]{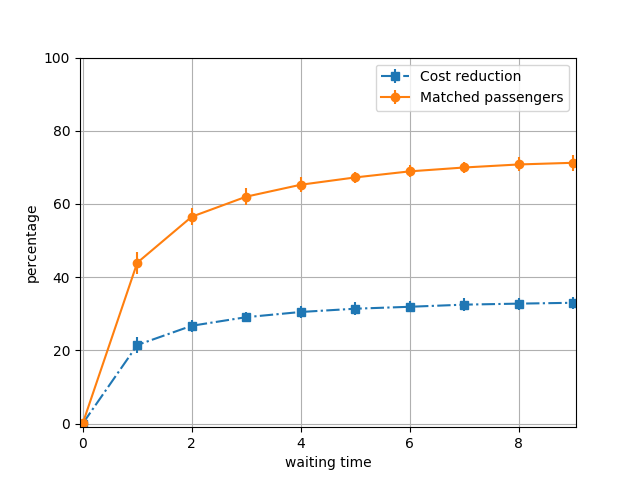} 
	\caption{Constant waiting time}
	\label{fig:subim7}
	\label{fig}
\end{figure}

In figure \ref{fig:subim7}constant waiting time assigned to all the passengers to compare our method for estimating the best waiting time with constant waiting time. Our results are close to two minutes constant waiting time and we achieve similar results with 1 minutes and 31 seconds as the average of our waiting time.

\section{Related Work}\label{Section 4}
Online ridesharing defines a scheme in which passengers arrive into the system as time goes on. At each time instance, only the information related to active passengers are known. Surveys of online ridesharing problem can be found in  \cite{agatz2011dynamic}, \cite{agatz2012optimization} presenting optimization challenges and \cite{furuhata2013ridesharing} addressing a classification of the current ridesharing systems.

Online ridesharing is in contrast with the case when entire data is available at the very beginning \cite{bernstein1993roommates}. In the traditional systems passengers have fixed schedules, starting points, and destinations and basically, all the passenger's information is available at any time instance \cite{michalak1994assessing}. Our problem investigates practical situations in the real world, and it is more challenging, as information emerges continuously and passengers must be matched optimally in a short notice. The automated ridesharing operator matches up potential passengers in a short notice.

This problem is very similar to pick up and delivery problems described in \cite{berbeglia2010dynamic}. It can be referred to like a particular case of dial-a-ride problem  \cite{attanasio2004parallel}. DARP mainly revolves around routing vehicles subject to time constraints on pickup and delivery. The majority of research in this category is carried out on static DARP where all the information is ready before initiating the system. These portions use two-phase scheduling strategy \cite{attanasio2004parallel, horn2002fleet} to route the taxi fleet. This is distinguished from our problem as we focus on passengers but in DARP scheduling taxis are considered as the main concerns. Moreover, here we aim to maximize matching of passengers while optimizing overall overheads, an approach that taxi-sharing problem ignores.

Li et al. \cite{li2014share} considered a problem in which a shared vehicle delivers a passenger and a parcel to destinations while persons have priority over parcels. They avoid serving two passengers simultaneously in a taxi. They present an exact MILP formulation to solve this problem in tiny instances to maximize taxi provider's profit in a deterministic configuration. They proposed a solution based on large neighborhood search heuristic for a case in which travel times and delivery locations are stochastic \cite{li2016share}. 

Ma et al. \cite{ma2013t} defined an algorithm to dispatch the best vehicle in a taxi sharing scheme when passengers send real-time queries. They provided a searching algorithm to schedule taxis to satisfy queries incurring minimum additional travel time. They showed using shared rides, it is possible to increase vehicle occupancy rate (25\% in their paper) having a relatively small increase in travel distances (only 13\%). Our problem differs from this research as beyond having real-time requests from passengers, we consider maximizing ridesharing partners and minimizing overheads. Taxi sharing problem (i.e., DARP, SARP) usually discussed regarding a scheduling problem, rather than a matching problem. We propose our problem in the sense of online roommate problem and online matching problem on general graphs \cite{bernstein1993roommates}.

Tao \cite{tao2007dynamic} addressed the problem where a set of passengers travel from different origins to a fixed destination on a regular basis and then dispatched from this fixed place to their origins in the reverse paths. The matching is done in real-time. A call center matches the closest taxi to the predetermined matching list of passengers. We also emphasize on the role of central coordinator to match passengers in a way the total cost of a fleet is minimized. However, passengers and matching sets are at the center of focus, not routing vehicles. Also, in our paper, passengers have distinguished starting points, and destinations and vehicle serve at most two passengers simultaneously.

\cite{herbawi2012ridematching} presented a genetic algorithm to solve the multi-criteria optimization model of minimizing total distance covered by drivers and incurred travel time of passengers while maximizing the size of matching. Detours of drivers and passengers time limitations are predetermined. Conversely, we calculate the optimal waiting time for every passenger such that resulting matching leads to optimal overall overhead. Additionally, we assume that passengers do not use a private vehicle to travel.

Agatz et al. \cite{agatz2011dynamic} designed optimization approaches to solve real-time matching in an online ridesharing scheme. They showed that using sophisticated optimization techniques benefits the performance of ridesharing systems. However, they opt for the case where passengers use their own vehicles to move. In our paper, we consider a limited fleet of independent taxis controlled by a central coordinator. In our paper, when the passengers start their trip they leave the system and will not participate in matching. But in \cite{ghoseiri2011real} riders can be matched while en-route.

Notwithstanding the fact that many ridesharing projects have been initiated in the past few years, many of them have been proved to be unsuccessful in practice \cite{ghoseiri2011real}. A ridesharing system must be flexible, efficient and economical to be substituted with the private cars having the advantage of immediate door-to-door access \cite{agatz2011dynamic}. Our model provides this features to passengers.

While the role of technological advances like GIS and communication networks and the potential of ridesharing are vastly studied \cite{sarraino2008spatial}, \cite{amey2011proposed}, developing sophisticated optimization and matching algorithms to arrange ridesharing partnerships in real-time has been largely neglected. In this paper, we mainly focus on designing an algorithm to maximize the rate of participation in ridesharing (matching passengers) while minimizing system-wide overheads.

A mixed integer optimization is developed by \cite{ghoseiri2011real} to maximize the matching of partners while considering individual preferences. In our paper, minimizing costs is a vital part of the objective function. waiting time is a predetermined factor manually selected by riders, but in our paper, it is computed optimally aligned with our perspective.

Some people may opt for the ride sharing merely for cost savings. Individuals not owning private cars have limited access to public transportation options can be other targets of ridesharing. People are likely to participate in ridesharing arrangements if they pay less fare in comparison with taking a taxi alone. Also, they have to accept a deviation from their shortest path too. There are different methods to share the travel-expenses among entities. The most straightforward intuitive way is to devide the costs evenly between passengers \cite{geisberger2009fast}. This approach would not be fair as the portion of the shared path could be different. To overcome this, passengers may divide the related costs proportional to the distances they ride autonomously and the part they shared the car. 

Among the topics investigated in this field, most studies capture passengers time preference by denoting an accepted time window. Agatz et. al \cite{agatz2011dynamic} considers an acceptable period between earliest departure time and latest arrival time to the destination defined by every passenger. Some papers limit the actual travel time by defining a fixed threshold deviation in traveling time \cite{amey2011proposed} and a maximum ride time \cite{baldacci2004exact}. In contrast, we propose a strategy where the central coordinator computes and declares an optimal waiting time by the advent of a passenger leading to optimal overall cost. A passenger's of tendency to move along longer paths to get to his desired destination is the first element to compute his optimal waiting time.

\cite{xiang2008study} introduced a ridesharing scheme considering a decentralized agent based system. In this paper, passenger agents try to find a partner every two minutes among driver agents. They delineated that given a high number of drivers, travel time for a single passenger decrease. We face a problem where passengers use taxis to commute not their own vehicles. Additionally, there is no emphasize on optimizing in \cite{xiang2008study} research.
\section{Conclusion and Outlook}
\label{section 5}
In this work, we have modeled a particular online ridesharing problem as a matching problem and provided an algorithm to solve it optimally. The case where passengers only use taxis and no one drives his private car is considered in this paper.
In addition, we have engineered NYC Taxis \& Limousine Commission dataset for testing the performance of our algorithm and predict real-time data.  By using our algorithm, we have shown that maximized participation in ridesharing arrangements decreases the total distance covered by taxis and consequently minimizes total costs in comparison with the case all passengers take taxis autonomously.
This is in favor of the central taxi provider since this scheme decreases the likelihood of losing requests as a result of lack of sufficient number of fleet.
There are many directions to extend this work. One is to implement other matching and optimization techniques to solve the same problem. Another direction is to extend the capabilities the algorithm to use the full capacity of the taxis by adjoining more passengers. One interesting research venue is to make the algorithm more flexible by enabling en route matchings. These initiatives can lead to more realistic scenarios.
\bibliographystyle{apalike}
\bibliography{Library.bib}
\end{document}